# Reply to Comment on
# "Six-body bound system calculations in the case of effective $\alpha$-core structure"

E. Ahmadi Pouya[1] and A. A. Rajabi[2]

*Physics Department, Shahrood University of Technology, P. O. Box 3619995161-316, Shahrood, Iran*

## Abstract

The present paper is provided in response to the comment of M. R. Hadizadeh *et.al* on our original paper "Six-body bound system calculations in the case of effective $\alpha$-core structure" [Eur. Phys. J. Plus (2016) **131**: 240]. In this paper we have presented our clarifications on their arguments about of the accuracy of the procedure of the calculations. In this paper our arguments turn out to be very efficient mainly discussed from the author's misunderstanding of the issues discussed in the original paper. In fact, their comment aims to exaggeratedly show that our paper is poor, but our following statements demonstrate that their comment is misleading and consequently unacceptable.

## I. Introduction

In the original paper entitled "Six-body bound system calculations in the case of effective $\alpha$-core structure" we investigated the six-body system as the effective $\alpha$-core structure as a bound system [1]. After revisions and replies to the reviewers' miscellaneous comments, the manuscript was ultimately published in EPJP in 25 July 2016. However, as research is generative, every discovery brings about further questions. In the first step, we would like to express our gratitude for their meticulous scrutiny into the article. Next, we are comprehensively dedicated to present our reply to the comment in order to compromise any misunderstandings apparent within the comment:

## II. Formalism

The authors of the comment aim to exaggeratedly show that our paper is completely poor, but our following statements are the opposite of their suggested comment. The number of Yakubovski components is 2700 for six-body as a non-identical particle system. In the case of identical particle system, we should solve 5 coupled equations. Therefore, the Yakubovsky formalism for the six-nucleon bound system leads to a set of five coupled equations [2] which can be reduced to a two coupled ones, for the case of effective $\alpha$-core structure, namely the two loosely bound neutrons with respect to the regarded $\alpha$-core nucleons [1]. In the original paper [1] first we have discussed and demonstrated that the first two components in fig. 1 in Ref. [1] contain the relevant configurations of the effective $\alpha$-core structure and the other components will not be taken into account [1]. This argument is a novel method that is not mentioned in previous works, like Ref. [2]. So, we applied this new idea as an appropriate strategy to attack the calculations of this case, namely effective $\alpha$-core structure, whereas the achieved results are completely genuine and desirable.

1. According to derivation of $P_{34}$ equation (2.28) in Ref. [3] in a partial-wave evaluation of the integral terms, the integrations are dependent on the angle variables $(x)$ of the shifted momenta. Therefore the shifted momenta are dependent on the basic Jacobi momenta with angles between them. For example the $P_{34}$ are given as [2]:

$$\langle a'|P_{34}|a''\rangle = \frac{1}{2}\frac{\delta(a'_1 - a''_1)}{(a''_1)^2}\frac{\delta(a'_4 - a''_4)}{(a''_4)^2}\frac{\delta(a'_5 - a''_5)}{(a''_5)^2}\int_{-1}^{1}dx_{2'3'}\frac{\delta[a''_2 - \tilde{a}''_2]}{(a''_2)^2}\frac{\delta[a''_3 - \tilde{a}''_3]}{(a''_3)^2} \quad (1)$$

With

$$\tilde{a}''_2 = \left|\frac{1}{3}\boldsymbol{a}'_2 + \frac{8}{9}\boldsymbol{a}'_3\right|; \; \tilde{a}''_3 = \left|\boldsymbol{a}'_2 - \frac{1}{3}\boldsymbol{a}'_3\right|; x_{2'3'} = \cos(\boldsymbol{a}'_2, \boldsymbol{a}'_3). \quad (2)$$

Where $x_{2'3'}$ is a general form without depending on any coordinate system and before selecting a suitable coordinate system, how do we define the azimuthal and spherical angles? As we all know after selecting coordinate systems in sect. 4 of Ref. [2] the $\varphi_{45}$ and $\varphi_{42'}$ are equal with $\varphi_5$ and $\varphi_{2'}$, respectively, because the fourth Jacobi vector is restricted in *x-z* plane (Eq. (23) and (24) in [2]). So, the first comment in the formalism is completely unreasonable and the Eq. (20) and (22), before selecting the suitable coordinate system, are completely correct. Likewise, the Eq. (25) and (26) are completely correct (with accurate integrations and evaluations) and undoubtedly consistent with Eq. (20) and (22).

---

[1] E.Ahmady.ph@ut.ac.ir
[2] A.A.Rajabi@shahroodut.ac.ir

2. In fact, in the numerical calculations both vector $\boldsymbol{a}_2$ and $\boldsymbol{a}'_2$ restricted to be parallel. Therefore, the $x_{22'}$ and $X_{22'}$ are equal to 1, without any dependence on angles.

3. This comment is obviously so exaggerated, because only the dot product was mistyped, so the Eqs. (23) and (24) without dot is completely correct.

4. and 9. : In the calculations we have considered $|\boldsymbol{a}''_2| = |\boldsymbol{a}'_2|$. Therefore, the Eqs. (A.9) and (A.10) is equal $|\pi_1| = |\pi_2|$, and also the $g(\pi_i)$ are scalars, therefore exchanging the label 1 and 2 is redundant, so this comment is inopportune.

5. , 6. , 7. 8. and 10. : The normalization factors are defined as our formalism, namely in the calculations we restricted the $\int d\varphi_{42'} \equiv 1$ not $2\pi$. So, these suggested coefficients are already concealed to the formalism.

11. This comment is also repetitive like first comment (1), and we have previously described it.

12. In $\tilde{a}_5^* = \left|\boldsymbol{a}'_2 - \frac{1}{3}\boldsymbol{a}'_3 - \boldsymbol{a}'_4 + \frac{1}{4}\boldsymbol{a}'_5\right|$, just "–" should be replaced with "+".

13. Eq. (A.23) is $\tilde{b}_2 = \left|\frac{1}{3}\boldsymbol{a}'_2 - \boldsymbol{a}'_3\frac{2}{3}\right|$, just 1/3 has been mistyped 1/2.

14. This comment is repetitive like 1 and 10.

15. The $\bar{a}_2 = \left|\frac{2}{3}\boldsymbol{b}'_2 - \frac{2}{3}\boldsymbol{b}'_3\right|$ is also scalar and the minus for plus sign is not effective in the formalism, considering the vertical position between two angles.

16. Such typos are not any effective in the main formalism and this formation is also correct, because the identity of these shifted momenta is correct.

In summary, this detailed reviews, confirms that our main formalism and standard derivation that is the main subject of the calculations are completely correct. Therefore, in our opinion their comment is so exaggerated and obviously unfriendly.

### III. Calculations

In our opinion in such numerical implementations the superfluous details of a typical calculation is not necessary and it is desirable to refer them. In addition, to verify the halo contributions of the two loosely bound neutrons, drawing the plots of Jacobi momenta, specially $\boldsymbol{a}_4$ and $\boldsymbol{b}_4$, is not currently accessible, because the total wave-function for effective $\alpha$-core structure system has 270 components, including 180 for $\psi_{12,123}^{1234}$ and 90 for $\psi_{12,12+34}^{1234}$ [2]. Also the momentum cutoff in our calculations is considered about 25 up 30 $MeV$. In order to verify the accuracy of our calculations and genuine results, we have again solved the coupled Yakubovsky integral equations, of course by previous codes applied in [1] and the codes have the same accuracy of the results for binding energies in tables 1 and 2 in the original paper [1]. To this aim, and for sure about our genuine results and also for comparison our new obtained results with their codes, we give new Yamaguchi parameters in table 1 and calculated 4-body bound system with these new parameters. In addition, the 6-body binding energy results are processing (it takes extra computing time) and we will report them in other opportunities.

Table1. New two-case Yamauchi parameters as well as represented binding energy results for 4-body bound system with Yamaguchi potentials that have been defined in Eq. (30) of the original paper [1].

| Case 1 | $\lambda = 0.200\ fm^{-3}$ | $\beta = 1.15\ fm^{-1}$ | $E_4 = -51.19\ MeV$ |
|---|---|---|---|
| Case 2 | $\lambda = 0.230\ fm^{-3}$ | $\beta = 1.20\ fm^{-1}$ | $E_4 = -57.72\ MeV$ |

By this evidence and these considerations, we believe that their comment in both formalism and numerical implementation is completely radical. In addition, the new obtained results that have represented in table 1, is quite enough to ensure us and them that the original paper has been reported genuine and desirable results.